\documentclass[nojss]{jss}

\usepackage{xcolor}

\author{Simon A. C. Taylor\\Lancaster University \And Timothy Park\\Lancaster University \And Idris A. Eckley\\Lancaster University}
\title{Multivariate Locally Stationary Wavelet Analysis with the \pkg{mvLSW} \proglang{R} Package}

\Plainauthor{Simon A. C. Taylor, Timothy Park, Idris A. Eckley} 
\Plaintitle{Multivariate Locally Stationary Wavelet Process Analysis with the mvLSW R Package} 
\Shorttitle{\pkg{mvLSW}: Multivariate Locally Stationary Wavelet Processes in R}

\Abstract{
This paper describes the \proglang{R} package \pkg{mvLSW}. The package contains a suite of tools for the analysis of multivariate locally stationary wavelet (LSW) time series. Key elements include: (i) the simulation of multivariate LSW time series for a given multivariate evolutionary wavelet spectrum (EWS); (ii) estimation of the time-dependent multivariate EWS for a given time series; (iii) estimation of the time-dependent coherence and partial coherence between time series channels; and, (iv) estimation of approximate confidence intervals for multivariate EWS estimates. A demonstration of the package is presented via both a simulated example and a case study with \code{EuStockMarkets} from the \pkg{datasets} package. This paper has been accepted by the Journal of Statistical Software. Presented code extracts demonstrating the \pkg{mvLSW} package is performed under version 1.2.1.}

\Plainkeywords{mvLSW, Multivariate evolutionary wavelet spectrum, coherence, R} 

\Address{
  Simon A. C. Taylor\\
  Department of Statistics and Mathematics\\
  Lancaster University\\
  Lancaster\\
  United Kingdom\\
  E-mail: \email{s.taylor2@lancaster.ac.uk}\\

  Timothy Park\\
  Department of Statistics and Mathematics\\
  Lancaster University\\
  Lancaster\\
  United Kingdom\\
  E-mail: \email{parkt@lancaster.ac.uk}\\

  Idris A. Eckley\\
  Department of Statistics and Mathematics\\
  Lancaster University\\
  Lancaster\\
  United Kingdom\\
  E-mail: \email{i.eckley@lancaster.ac.uk}\\
  URL: \URL{http://www.lancaster.ac.uk/~eckley}}

\usepackage{amsmath,amsfonts}
\usepackage{caption}
\usepackage{subcaption}
\newcommand{\Spec}{\mathbf{S}}
\newcommand{\V}{\mathbf{V}}
\newcommand{\I}{\mathbf{I}}

\begin{document}
\section[Introduction]{Introduction}\label{sec:introduction}

Technological advances in sensors and other data recording mechanisms have led to an increased need to efficiently and accurately analyse multivariate time series. Areas of research where such methods are commonly required include economics \citep{Rua09}, medicine \citep{Cribben12}, telecommunications \citep{Bardwell16} and environmental science \citep{Sha07}. Historically, one might approach such a multivariate time series challenge by making the assumption that the underlying process was second order stationary. However, there are an increasing number of cases where such global stationarity assumptions are not tenable due to the second order structure of the process changing over time. Such series are said to be non-stationary.

Over the years, a number of popular approaches for modelling non-stationary time series have been proposed, though activity has predominantly focussed on the univariate time series setting. Notable contributions in the univariate setting include the seminal work of Priestley on evolutionary processes \citep{Priestley65}, time varying moving average methods \citep{Hallin86} and locally stationary processes \citep{Dahlhaus1997, Nas00}. For a recent review of the literature in this area, we refer readers to the excellent review article by \citet{Dah12}. Implementations of these models in \proglang{R} \citep{RTeam15} include the \pkg{LSTS} \citep{Ole15_RLSTS} and \pkg{wavethresh} \citep{Nas08, Nas13_Rwavethresh}. These respectively implement the locally stationary Fourier and wavelet approaches for univariate time series. A multidimensional implementation of a locally stationary framework is also available for regular lattice processes (see \citet{ENT2010} and \citet{EN2011} for details). 
 
We are, of course, by no means the first to consider the challenge of modelling non-stationary multivariate time series. Early contributions include those by \citet{Omb05} who extend the smoothed localised exponential (SLEX) model to the multivariate setting and \citet{San10} who apply the wavelet methodology to bivariate and multivariate time series respectively. The SLEX framework, originally proposed by \citet{Omb02}, seeks to segment a non-stationary time series into dyadic stationary blocks that then permit the employment of standard methods to form a time-dependent Fourier analysis. The multivariate extension of SLEX, proposed by \citet{Omb05}, enables the estimation of the cross-spectrum and investigations into the coherence structure. Conversely, the work of \citet{San10} and \citet{Cho15} decompose the dependence within a locally stationary time series into two multiscale components; the within-channel spectral structure and the cross-spectrum (between-channel) structure for the bivariate and p-variate time series cases respectively. However neither of these recent wavelet-based contributions directly address a key modelling challenge for truly multivariate non-stationary signals, namely the identification of whether whether the connection between two channels is either direct or indirect (i.e., driven by other observed channel(s)). 

This article presents the \proglang{R} implementation of an alternative formulation for multivariate locally stationary wavelet (LSW) model, proposed by \citet{Par14}. The work extends the locally stationary wavelet framework of \citet{Nas00} to a multivariate setting, also enabling the estimation of the within-channel spectral structure and cross-spectrum. \citet{Par14} also introduce the concepts of local coherence and, crucially, local partial coherence within the multivariate locally stationary wavelet setting. 
 The \pkg{mvLSW} package \citep{mvLSWpkg} implements the work of \citet{Par14}, building upon the univariate locally stationary wavelet time series implementation in \pkg{wavethresh} \citep{Nas08, Nas13_Rwavethresh}. Specifically \pkg{mvLSW} provides functionality for: the simulation of multivariate LSW time series for a given evolutionary wavelet spectrum (EWS); estimation of the multivariate EWS for a given time series with point-wise confidence intervals; and, estimation of the local coherence and local partial coherence between time series channels. The \pkg{mvLSW} package is available for download from the comprehensive \proglang{R} archive network (CRAN). Note that throughout this article all \pkg{mvLSW} package references relate to version 1.2.1 of the package, as available on CRAN.

The paper is structured as follows. The framework of multivariate LSW modelling is briefly introduced in Section~\ref{sec:framework} together with the introduction of the \pkg{mvLSW} package for defining a multivariate EWS and simulating a multivariate LSW time series. Section~\ref{sec:mvEWS} describes estimation of the multivariate EWS and its implementation in \proglang{R}. The concept and estimation of localised coherence and partial coherence is discussed in Section~\ref{sec:coh}. We then introduce the functions developed to implement this coherence estimation framework, prior to summarising the mechanism by which approximate confidence intervals can be constructed in Section~\ref{sec:CI}. A complete demonstration of the \pkg{mvLSW} package is presented in Section~\ref{sec:CS_EU} including a case study which makes use of the European financial index time series \code{EuStockMarkets} that is accessible from the \pkg{datasets} package. 

$~$\\
$~$\\

\section[Framework]{Framework} \label{sec:framework}

This section briefly introduces the multivariate locally stationary wavelet (LSW) time series and introduces its implementation within \proglang{R} in the \pkg{mvLSW} package. Section \ref{sec:mvLSW} introduces the multivariate LSW framework of \citet{Par14} and a description of the time-frequency power decomposition by the multivariate evolutionary wavelet spectrum (EWS). Section~\ref{sec:EG} presents a worked example for a trivariate LSW process and demonstrates how to simulate a time series with the specified spectral form.

\subsection{The multivariate LSW process}\label{sec:mvLSW}

Adopting the notation of \citet{Par14}, the doubly indexed $P$-variate stochastic process $\{\mathbf{X}_{t;T}\}$ for $t=1,\ldots,T$ with dyadic length, i.e., $T=2^J$ for some $J\in\mathbb{N}$, is said to be a multivariate LSW process if it is represented by:
\begin{eqnarray}
  \mathbf{X}_{t;T} & = & \sum_{j=1}^{\infty}\sum_{k} \V_j(k/T) \psi_{j,k}(t) \mathbf{z}_{j,k}.\label{eq:Model}
\end{eqnarray}
As in the earlier univariate LSW work of \citet{Nas00}, the $\{\psi_{j,k}\}$ denote the set of discrete non-decimated wavelets for each level $j$ and location $k$ pair such as the Daubechies compactly supported wavelet \citep{Dau90}. The random vectors $\{\mathbf{z}_{j,k} = (z_{j,k}^{(1)}, \ldots, z_{j,k}^{(P)})^{\top}\}$ are uncorrelated innovations with zero expectation and $P{\times}P$ identity variance-covariance matrix. Finally, $\V_j(u)$ denotes the lower-triangular $P{\times}P$ transfer function matrix for the given level index at rescaled time $u:=t/T\in(0,1)$ . Some smoothness conditions are assumed for the elements of the transfer function matrix, controlling its behaviour \citep[see][for details]{Par14}. For notational ease, the explicit dependence on $T$ shall henceforth be suppressed but its dependence on the process and derived estimates shall naturally be assumed.

Within the above modelling framework, arguably the key element of interest is the transfer function matrix. Since the innovations are orthogonal with unit variance and the wavelets are not channel specific, then all forms of dependency between and within channels must be encapsulated by the transfer function matrix. For instance, the diagonal element $V_j(u)^{(p,p)}$ for channel $p$, for $p = 1, \ldots, P$, defines the time-varying amplitudes of the respective wavelet at level $j$ akin to the univariate scenario of Definition~1 in \citet{Nas00}. Furthermore, if the off-diagonal element $V_j(u)^{(p,q)}$, for $p>q$, is non-zero then there is a time-varying dependence between channels $p$ and $q$, but if this element is zero for all time and levels then the two channels are uncorrelated.

The transfer function matrix is also useful in forming the spectral representation of a multivariate LSW time series. Specifically, the power contained within a multivariate LSW process is described by the multivariate evolutionary wavelet spectrum (EWS) taking the form:
\begin{eqnarray}
  \Spec_j(u) & = & \V^\top_j(u) \V_j(u),\label{eq:mvEWS}
\end{eqnarray}
where $\V^\top_j(u)$ denotes the transpose of $\V_j(u)$. The diagonal element $S^{(p,p)}_j(u)$ denotes the auto-spectrum for channel $p$, whilst the off-diagonal element $S^{(p,q)}_j(u)$, for $p \neq q$, denotes the cross-spectrum between the channel pair $p$ and $q$, and it describes how the power in the process is shared between channels.

\subsection[Trivariate worked example]{Trivariate worked example} \label{sec:EG}

The multivariate EWS defined in Equation~\ref{eq:mvEWS} is a concise description of the time-varying distribution of power for a multivariate LSW time series. To demonstrate this, we introduce a trivariate LSW process which will be used as a worked example throughout the paper. We assume that the process is built using discrete Haar wavelets, and that the EWS for this process has non-zero power at only the second finest level, $j=2$, given by:
\begin{eqnarray}
  \Spec_2(u) & = & \left[\begin{array}{ccc}
    4+16u & 2+8u & 2+8u   \\
    2+8u  & 6    & 1+4u   \\
    2+8u  & 1+4u & 20-14u \\
  \end{array}\right], \quad \mbox{for}~u \in (0,1). \label{eq:S2_eg}
\end{eqnarray} 

Note that the auto-spectrum for the second channel is constant and so this channel is marginally stationary. The auto-spectrum for the first channel increases over the unit time interval and so, with the spectrum at all other levels equal to zero, the variability of this channel steadily increases. Conversely, the auto-spectrum for the third channel decreases over time. Dependence between the channels is also time-varying within this example.

The following extract defines the trivariate EWS in \proglang{R} using the \pkg{mvLSW} package for a time series of length $T=1024$ (i.e., $J=10$):
\begin{CodeChunk}
\begin{CodeInput}
R> library("mvLSW")
R> P <- 3        ## Number of channels
R> T <- 1024     ## Time series length
R> J <- log2(T)  ## Number of levels
R> Spec <- array(0, dim = c(P, P, J, T))
R> Spec[1, 1, 2, ] <- seq(from = 4, to = 20, length = T)
R> Spec[2, 2, 2, ] <- 6
R> Spec[3, 3, 2, ] <- seq(from = 20, to = 6, length = T)
R> Spec[1, 2, 2, ] <- Spec[2, 1, 2, ] <- seq(from = 2, to = 10, len = T)
R> Spec[1, 3, 2, ] <- Spec[3, 1, 2, ] <- seq(from = 2, to = 10, len = T)
R> Spec[2, 3, 2, ] <- Spec[3, 2, 2, ] <- seq(from = 1, to = 5, len = T)
R> True_mvEWS <- as.mvLSW(x = Spec, filter.number = 1, 
+    family = "DaubExPhase", names = c("X1","X2","X3"))
\end{CodeInput}
\end{CodeChunk}

The command \code{as.mvLSW} converts a 4D array into an object having \proglang{S3} class \code{mvLSW}. This class is central to the \pkg{mvLSW} package and consists of two components: the multivariate EWS as a 4D numerical array and a list of sundry information about the spectrum such as its dimensions, analysing wavelet and channel names. The input arguments for \code{as.mvLSW} are:

\begin{itemize}
  \item \code{x}: The multivariate EWS as a 4D array (${P}\times{P}\times{J}\times{T}$).
  \item \code{filter.number}, \code{family}: analysing wavelet \citep[see][]{Nas13_Rwavethresh}.
  \item \code{names}: Channel names (optional).
\end{itemize}

\begin{figure}
  \centering
  \includegraphics[width=0.7\textwidth]{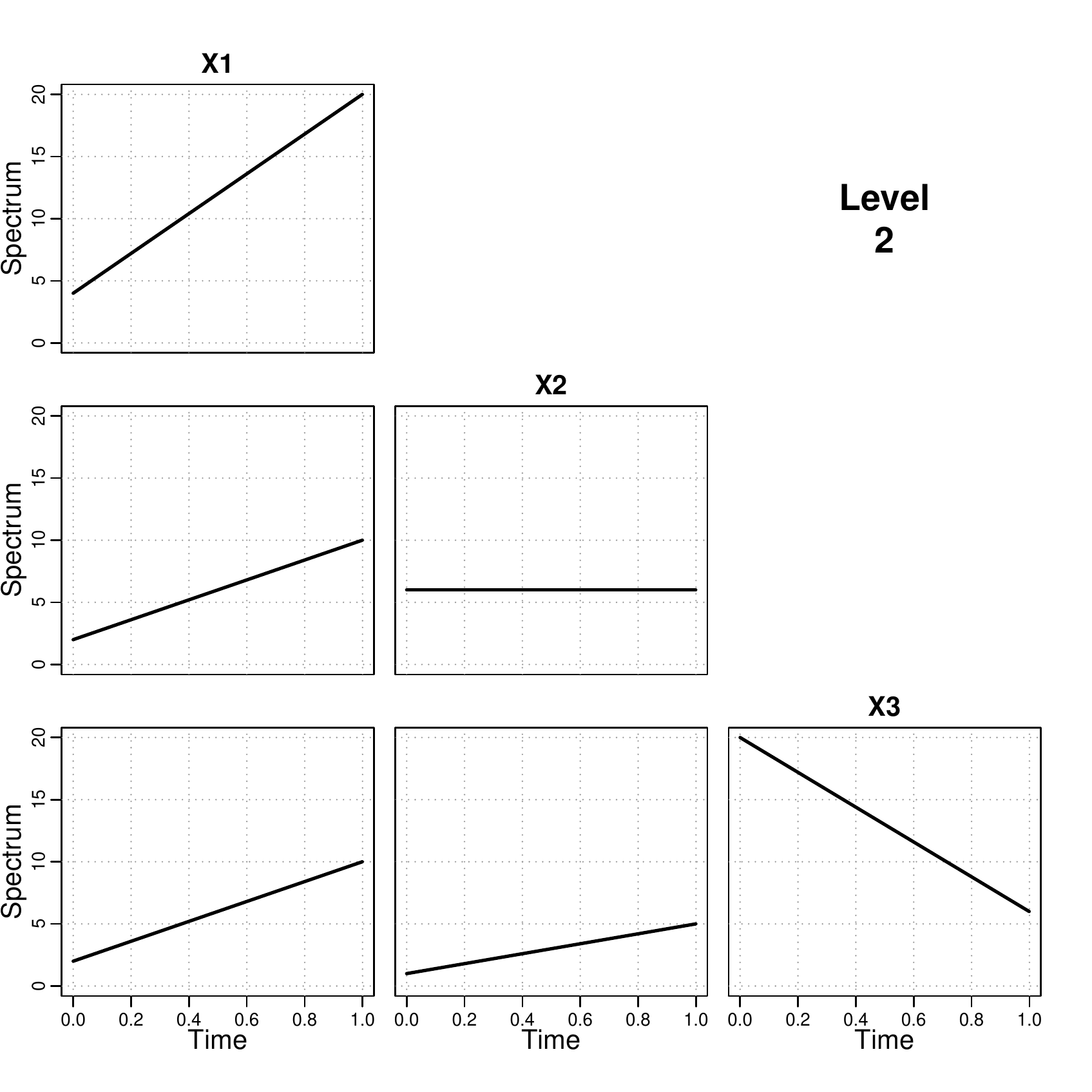}
  \captionof{figure}{True trivariate EWS at level 2 defined by Equation~\ref{eq:S2_eg}.}
  \label{fig:True_Spec}
\end{figure}

Due to the dimensions of a multivariate EWS, visualisations that illustrate the distribution of power over levels and locations can only be obtained via slices through this 4D object. For instance, the plot in Figure~\ref{fig:True_Spec} displays the trivariate EWS at the second level defined in Equation~\ref{eq:S2_eg} that is invoked by the following command:
\begin{CodeChunk}
\begin{CodeInput}
R> plot(True_mvEWS, style = 2, info = 2, ylim = c(0, 20), lwd = 2)
\end{CodeInput}
\end{CodeChunk}

The essential input arguments used here are:
\begin{itemize}
  \item \code{x}: An object of \code{mvLSW} class.
  \item \code{style}: Numerical index defining the plotting style:
  \begin{enumerate}
    \item Single spectra plot between channel pair $(p, q)$ and at level $j$ is specified via \code{info = c(p, q, j)}.
    \item Panel plot across channels at level $j$ is specified via \code{info = j}.
    \item Panel plot across levels for channel pair $(p, q)$ is specified via \code{info = c(p, q)},
    \item As \code{style = 3}, but presented as an image plot.
  \end{enumerate}
\end{itemize}

Additional arguments such as \code{ylim} and \code{lwd} can be supplied to customise the generated plot. Point-wise approximate confidence intervals introduced in Section~\ref{sec:CI} can be drawn by supplying a list containing the upper and lower bounds via the optional \code{Interval} argument. Furthermore, the inclusion of the logical argument \code{diag}, when \code{style = 2}, suppresses the drawing of the diagonal plots on the panel. This is particularly useful for plotting the local coherence and partial coherence estimates discussed in Section~\ref{sec:coh}.

The transfer function matrix is obtained by factorising the multivariate EWS matrix using Cholesky decomposition. A realisation of a multivariate LSW process with a pre-specified EWS is therefore simulated using Equation~\ref{eq:Model} with innovations sampled independently from some defined distribution that has zero mean and unit variance. This procedure is implemented by the \code{rmvLSW} command. For example, the following extract generates a realisation of a trivariate LSW time series, with EWS structure as defined by Equation~\ref{eq:S2_eg} with Gaussian innovations:
\begin{CodeChunk}
\begin{CodeInput}
R> set.seed(1)
R> X <- rmvLSW(Spectrum = True_mvEWS, noiseFN = rnorm)
R> plot(x = X, main = "Gaussian mvLSW time series")
\end{CodeInput}
\end{CodeChunk}

\begin{figure}
  \centering
  \includegraphics{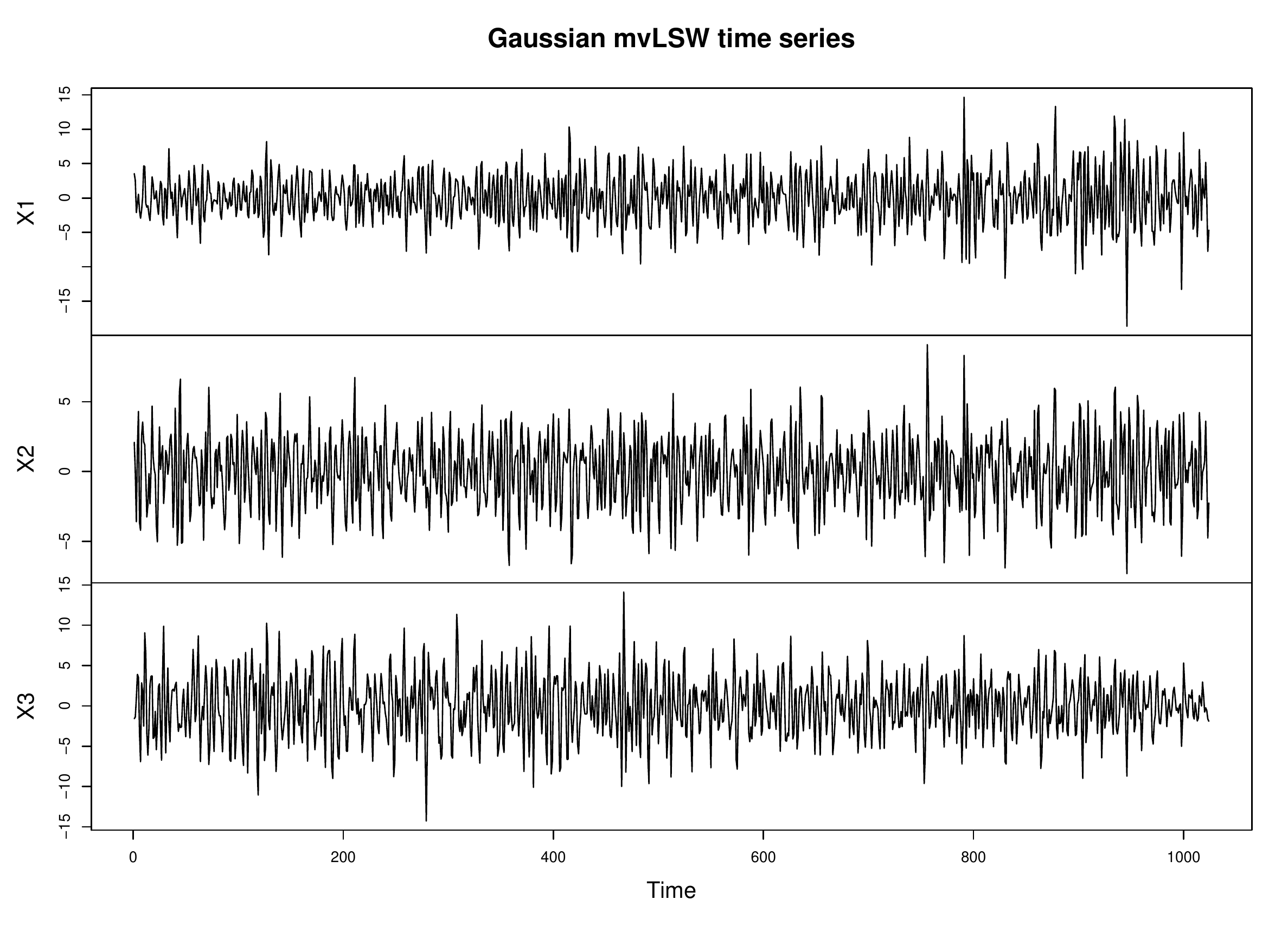}
  \caption{A realisation of a (Gaussian) trivariate locally stationary wavelet process having EWS structure as defined in Equation~\ref{eq:S2_eg}.}
  \label{fig:Sample_TimeSeries}
\end{figure}

The input arguments for \code{rmvLSW} are:
\begin{itemize}
  \item \code{Spectrum}: The multivariate EWS as a \code{mvLSW} object.
  \item \code{noiseFN}: Function for generating the orthogonal innovation process with zero expectation and unit variance. Additional arguments provided to \code{rmvLSW} are passes to the supplied function.
\end{itemize}
The \proglang{S3} \code{simulate} method can alternatively be called to simulate a realisation of a multivariate LSW time series for a provided multivariate EWS. An example of simulated trivariate LSW time series is presented in Figure~\ref{fig:Sample_TimeSeries}.

$~$\\
$~$\\

\section[Multivariate EWS estimation]{Multivariate EWS estimation}\label{sec:mvEWS}

The multivariate LSW framework presented in the previous section enables the generation of a time series that possess spectral properties defined by a specified multivariate EWS. Conversely, interest often lies in understanding the spectral structure for a given multivariate LSW time series. Such a scheme was proposed by \citet{Par14}, estimating the multivariate EWS, $\{\hat{\Spec}_{j,k}\}$ for levels $j = 1, \ldots, J$ and locations $k = 0, \ldots, T-1$. To begin, each channel of the time series is independently transformed based on a given wavelet to determine the empirical wavelet coefficient vector, $\mathbf{d}_{j,k}=[d^{(1)}_{j,k},\ldots,d^{(P)}_{j,k}]^\top$, with elements:
\begin{eqnarray}
  d^{(p)}_{j,k} & = & \sum_{t=1}^{T} X_{t}^{(p)} \psi_{j,k}(t).\label{eq:d}
\end{eqnarray}
The wavelet periodogram is then defined by the matrix set for each level and location pair as:
\begin{eqnarray}
  \I_{j,k} = \mathbf{d}_{j,k} \mathbf{d}^\top_{j,k}. \label{eq:RawPeriod}
\end{eqnarray}

Mirroring a result of \citet{Nas00} for univariate time series, \citet{Par14} established that Equation~\ref{eq:RawPeriod} is both biased and has non-vanishing estimator variance. These issues are overcome by smoothing and correcting the raw periodogram estimate in defining the multivariate EWS estimator:
\begin{eqnarray}
  \hat{\Spec}_{j,k} = \sum_{l=1}^J \sum_{m=-M}^M (\mathrm{A}^{-1})_{j,l}~ w_M(m) \I_{l,k+m}. \label{eq:S_hat}
\end{eqnarray}
Here, $w_M(m)$ denotes the symmetric kernel function on the compact support $m\in[-M, M]$ \citep{Omb01a} and $(\mathbf{A}^{-1})_{j,l}$ is the $(j,l)$-element from the inverted autocorrelation wavelet inner product matrix \citep{Eck05}.

Perhaps unsurprisingly, the bias correction in Equation~\ref{eq:S_hat} means that the spectral matrix estimates $\{\hat{\Spec}_{j,k}\}$ are no longer guaranteed to be positive definite. This leads to difficulties in evaluating quantities that are derived from the multivariate EWS, such as the localised coherence and partial coherence which we introduce in Section~\ref{sec:coh}. The matrices of the multivariate EWS estimate must therefore be regularized using, for example, the approach of \citet{Sch99} to ensure that subsequently derived estimates are themselves valid.

\subsection[Trivariate worked example continued]{Trivariate worked example continued} \label{sec:EWS_EG}

Estimation of the multivariate EWS is implemented using the \code{mvEWS} function. The following code extract estimates the trivariate EWS for the time series simulated in Section~\ref{sec:EG}. The wavelet transform is performed using the Haar wavelet and smoothing of the periodogram is performed using the rectangular kernel $w_M(m) = (2M+1)^{-1}$ for $m\in[-M,M]$ with parameter $M=\sqrt{T}=32$.
\begin{CodeChunk}
\begin{CodeInput}
R> EWS <- mvEWS(X = X, filter.number = 1, family = "DaubExPhase",
+    kernel.name = "daniell", kernel.param = sqrt(T), bias.correct = TRUE,
+    tol = 1e-10)

\end{CodeInput}
\end{CodeChunk}
The arguments of \code{mvEWS} are:
\begin{itemize}
  \item \code{X}: A regular $P$-variate time series with dyadic length of class \code{matrix} or \code{ts}, \code{xts} \citep{Rya14_Rxts} or \code{zoo} \citep{Zei16_Rzoo}.
  \item \code{filter.number}, \code{family}: the wavelet, passed to \code{wd} from \pkg{wavethresh}.
  \item \code{kernel.name}, \code{kernel.param}: Defines the smoothing kernel that is evaluated by the base command \code{kernel}. The argument \code{kernel.name} is a character string that names any kernel type evaluated by \code{kernel}.
  \item \code{bias.correct}: A logical variable, indicating whether the biased or corrected estimator should be estimated.
  \item \code{tol}: Threshold applied in matrix regularization.
\end{itemize}

Estimation of the trivariate EWS involves a number of steps including wavelet transformation, smoothing, bias correction and regularisation. The function returns an object of class \code{mvLSW}. Details about each step are stored within this object, under the information list. A summary of these details can be examined by invoking the \code{print} or \code{summary} command.
\begin{CodeChunk}
\begin{CodeInput}
R> summary(EWS)
\end{CodeInput}
\begin{CodeOutput}
== Dimensions ==
P : 3 
J : 10 
T : 1024 

== Wavelet Transform ==
Family        : DaubExPhase 
Filter Number : 1 

== Smoothing ==
Type   : all
Method : Daniell(32) - GCV criterion = 38.68027

== Sundries ==
Applied Bias Correction : TRUE 
Minimum Eigenvalue      : 1e-10 

\end{CodeOutput}
\end{CodeChunk}
The printed summary states the dimension of the object, the wavelet and smoothing method used in deriving the estimate. The item \code{Type} under the smoothing heading identifies that the same kernel function is applied to all levels. Smoothing can be performed on a by-level basis. For further details, the reader is referred to the package's manual. A measure of the smoothing performance is calculated using a generalized cross validation (GCV) criterion \citep{Omb01a}. The sundries information states whether or not the bias correction has been applied and the minimum eigenvalue across all time and level pairs to confirm whether all estimated spectral matrices are positive definite. 

We should note that the package requires a time series of dyadic length. In practice, not all series will be of length $2^J$ for some $J \in \mathbb{N}$. As \citet{NS94} describe, various techniques can be used to overcome this challenge. These include the use of reflective or periodic boundary conditions; zero padding and truncating either the head or tail of the data. Naturally, there can be various advantages and disadvantages associated with each of these approaches, depending on the underlying data generation process \citep{Per00}.

$~$\\

\section[Localised coherence and partial coherence]{Localised coherence and partial coherence} \label{sec:coh}

Dependence within a multivariate time series can occur both within and between channels. Coherence provides a measure of the linear dependence between any channel pair. However, if the multivariate time series is non-stationary then the coherence measure may vary over time. \citet{Par14} introduced the concept of local coherence based on the locally stationary wavelet framework defined in Section~\ref{sec:mvLSW}. At a particular level, the local coherence matrix function is defined by:
\begin{eqnarray}
  \boldsymbol{\rho}_j(u) & = & \mathbf{D}_j(u) \Spec_j(u) \mathbf{D}_j(u),\label{eq:Coher}
\end{eqnarray}
where $\mathbf{D}_j(u) = \mathrm{diag}\{ [S^{(p,p)}_j(u)]^{-1/2} : p=1,\ldots,P \}$. The off-diagonal elements of the localised coherence matrices may take any value within $[-1,1]$ where strong linear dependence is identified by values near $\pm1$. Whilst this coherence measure provides an indication of strong dependence between a channel pair, it cannot distinguish whether this relationship is direct or indirect. In other words, it cannot identify whether the relationships occur because of direct dependencies with other (observed) channels. To quantify direct linear dependence, \citet{Par14} introduced the local partial coherence matrix function. For any given level, this is defined as follows:
\begin{eqnarray}
  \mathrm{\Gamma}_j(u) & = & -\mathbf{H}_j(u) \mathbf{G}_j(u) \mathbf{H}_j(u),\label{eq:PCoher}
\end{eqnarray}
where $\mathbf{G}_j = \Spec_j(u)^{-1}$ and $\mathbf{H}_j(u) = \mathrm{diag}\{ [G^{(p,p)}_j(u)]^{-1/2} : p=1,\ldots,P \}$.

\subsection[Trivariate worked example continued]{Trivariate worked example continued}\label{sec:coh_eg}

Within the \pkg{mvLSW} package both forms of coherence can be estimated using the \code{coherence} command. This implements a `plug-in' estimator proposed by \citet{Par14}. The input arguments for \code{coherence} are:
\begin{itemize}
  \item \code{object}: The multivariate EWS as a \code{mvLSW} object.
  \item \code{partial}: A logical variable, indicating whether to evaluate the partial coherence?
\end{itemize}
Here, the distinction between evaluating the localised coherence and localised partial coherence is simply achieved by toggling the logical argument as demonstrated below with the trivariate EWS estimate derived in Section~\ref{sec:EWS_EG}:
\begin{CodeChunk}
\begin{CodeInput}
R> RHO <- coherence(object = EWS, partial = FALSE)
R> GAMMA <- coherence(object = EWS, partial = TRUE)
\end{CodeInput}
\end{CodeChunk}

As with the multivariate EWS, the returned coherence measure is stored within a 4D array of class \code{mvLSW}. A visualisation of a slice through this object can be generated by invoking the \code{plot} command as described in Section~\ref{sec:EG} for appropriate \code{style} and \code{info} arguments. Note that the diagonal elements of both coherence matrices do not contain any useful information and so can be suppressed when generating the image by toggling the \code{diag} logical argument as demonstrated in the following extract for the trivariate example: 
\begin{CodeChunk}
\begin{CodeInput}
R> plot(x = RHO, style = 2, info = 2, diag = FALSE, ylim = c(-1, 1), 
+    lwd = 2, ylab = "Coherence")     
R> plot(x = GAMMA, style = 2, info = 2, diag = FALSE, ylim = c(-1, 1), 
+    lwd = 2, ylab = "P. Coh.")          
\end{CodeInput}
\end{CodeChunk}

\begin{figure}
 \centering
 \includegraphics[width=0.48\textwidth]{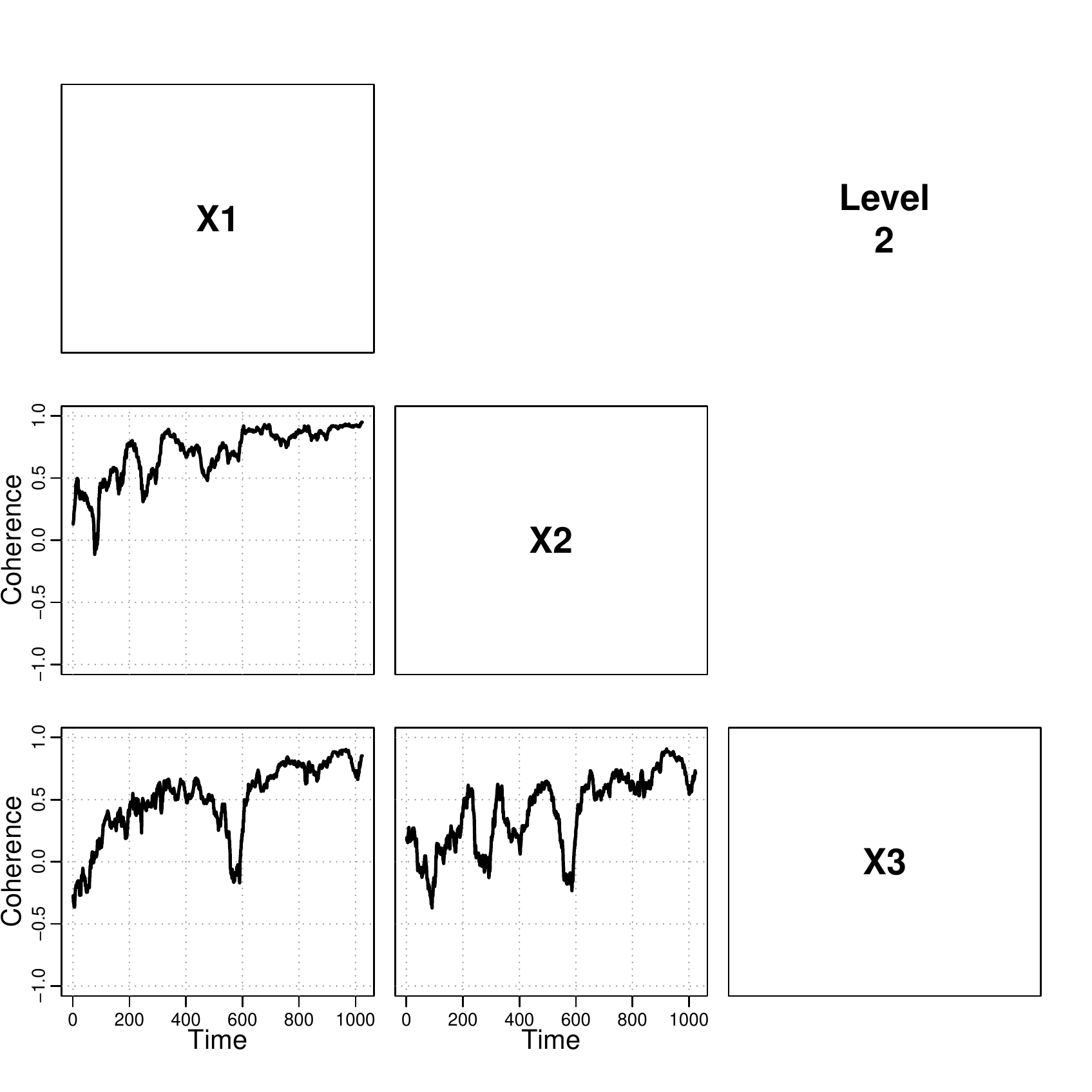}
 \includegraphics[width=0.48\textwidth]{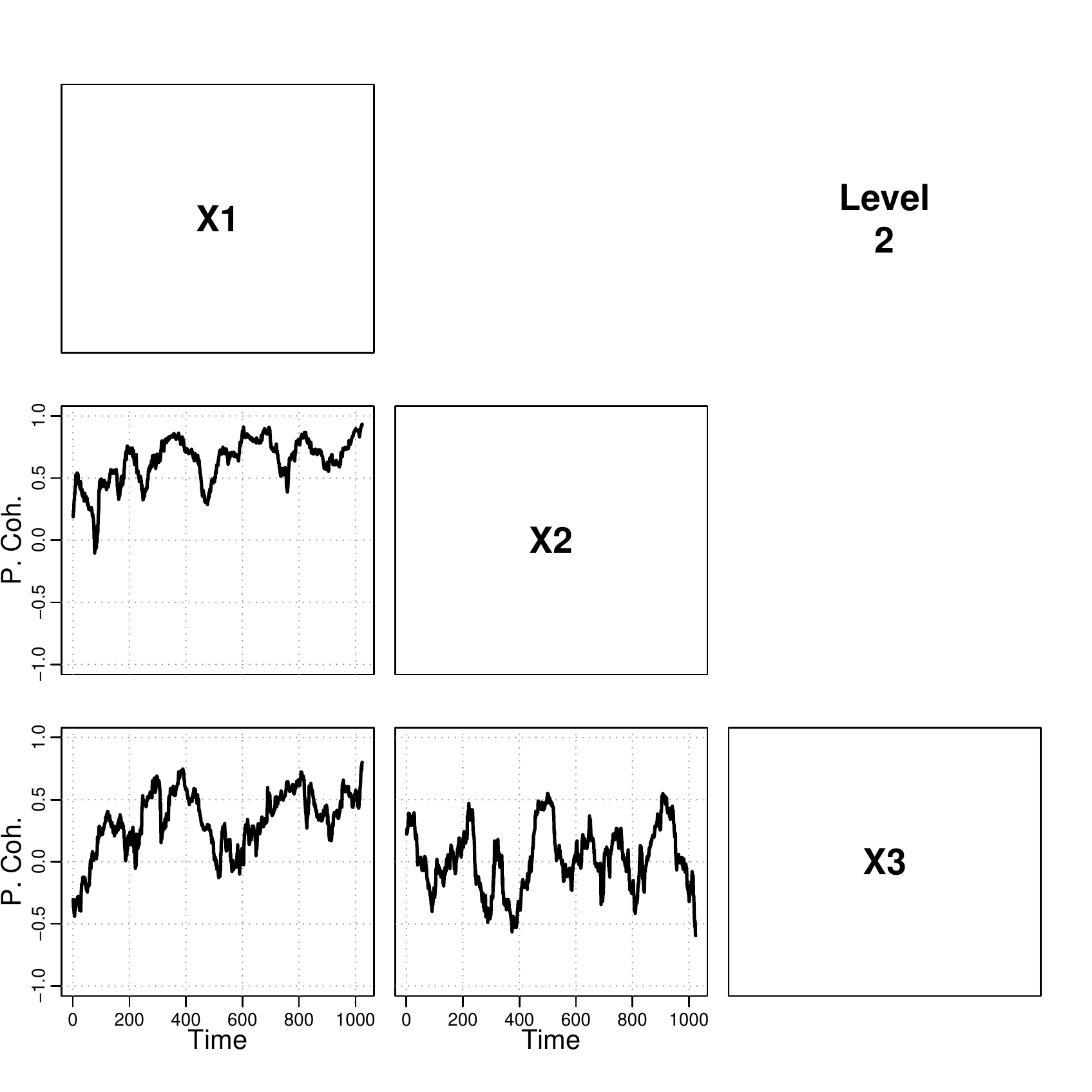}
 \caption{Localised coherence (left) and partial coherence (right) estimates at level 2 for the simulated trivariate time series.}
 \label{fig:Sample_RhoGam}
\end{figure}

The localised coherence and partial coherence estimates at the second level are presented in Figure~\ref{fig:Sample_RhoGam}. Recalling the form of the EWS in Equation~\ref{eq:S2_eg}, it is perhaps unsurprising that the estimates of the localised coherence at the second level are mostly positive and gradually increase over time. However, the localised partial coherence estimate between the second and third channels is closer to zero compared to its localised coherence estimate, suggesting that the measured linear dependence can mostly be explained by their dependence with the first channel.

$~$\\
$~$\\
$~$\\

\section[Confidence intervals for the multivariate EWS estimate]{Confidence intervals for the multivariate EWS estimate} \label{sec:CI}

Under reasonably mild assumptions, \citet{Par14b} established that the approximate, point-wise 100($1-\alpha$)\% confidence interval for a given element of the multivariate EWS, $S_{j,k}^{(p,q)}$, is given by:
\begin{eqnarray}
\left\{\hat{S}_{j,k}^{(p,q)} - \Phi\left(1-\frac{\alpha}{2}\right) \sqrt{\mathrm{Var}\left[\hat{S}_{j,k}^{(p,q)}\right]}, \quad \hat{S}_{j,k}^{(p,q)} + \Phi\left(1-\frac{\alpha}{2}\right)\sqrt{\mathrm{Var}\left[\hat{S}_{j,k}^{(p,q)}\right]}\right\}, \label{eq:mvEWS_CI}
\end{eqnarray}
where $\Phi(x)$ denotes the standard Gaussian cumulative distribution function. The asymptotic variance of an element of the multivariate EWS estimator is easily shown to be:
\begin{eqnarray}
  \mathrm{Var}\left[\hat{S}_{j,k}^{(p,q)}\right] = \sum_{l_1,l_2=1}^J ~ \sum_{m_1,m_2=k-M}^{k+M} (A^{-1})_{j,l_1} (A^{-1})_{j,l_2} ~ w_M(m_1) w_M(m_2) ~ \mathrm{Cov}\left[I_{l_1,m_1}^{(p,q)}, I_{l_2,m_2}^{(p,q)}\right], \label{eq:VarS_hat}
\end{eqnarray}
where the covariance between any pair of wavelet periodogram elements is:
\begin{eqnarray}
  \mathrm{Cov}\left[I_{j,k}^{(p,q)}, I_{l,m}^{(p,q)}\right] & = & \left[\sum_{h=1}^J B_{j,l,h}(m-k) S_h^{(p,q)}\left(\frac{k+m}{2T}\right)\right]^2 \nonumber\\ 
  && + \prod_{r=p,q}\sum_{h=1}^J B_{j,l,h}(m-k) S_h^{(r,r)}\left(\frac{k+m}{2T}\right) + \mathcal{O}\left(T^{-1}\right). \label{eq:CovII}
\end{eqnarray}

Here, $B_{j,l,h}(\lambda) = \sum_\tau \Psi_{j,h}(\tau)\Psi_{l,h}(\tau-\lambda)$ is the inner product of cross-level autocorrelation wavelet inner product function where $\Psi_{j,l}(\tau) = \sum_{k}\psi_{j,k}(0)\psi_{l,k+\tau}(0)$ \citep{Fry06}. Note that the special case $\{A_{j,l} = B_{j,j,l}(0)\}$ for levels $j,l = 1,\ldots, J$ denotes the inner product matrix that quantifies the leakage bias of the raw periodogram, see Section~\ref{sec:mvEWS}. The variance is dependent on knowing the true EWS but this is typically unavailable. Consequently, the variance is estimated by substituting the multivariate EWS estimate $\hat{S}_{j,k}$.

\subsection[Trivariate worked example continued]{Trivariate worked example continued}\label{sec:CI_eg}

Before calculating the asymptotic variance, the autocorrelation wavelet inner product function $B_{j,l,h}(\lambda)$ must first be calculated. The command \code{AutoCorrIP} evaluates the inner product functions for a given wavelet such as the Haar wavelet demonstrated below:
\begin{CodeChunk}
\begin{CodeInput}
R> HaarACWIP <- AutoCorrIP(J = J, filter.number = 1, family = "DaubExPhase")

\end{CodeInput}
\end{CodeChunk}
The \code{AutoCorrIP} command returns a $(2T+1)\times{J}\times{J}\times{J}$ array that corresponds to the lag $\lambda$ and level indices $j$, $l$ and $h$ respectively.

As with the localised coherence and partial coherence estimates, the variance is evaluated as a `plug-in' estimator based on a given multivariate EWS estimate. This is implemented by \code{varEWS} as demonstrated below for the trivariate EWS estimate evaluated in Section~\ref{sec:EWS_EG}:
\begin{CodeChunk}
\begin{CodeInput}
R> VAR <- varEWS(object = EWS, ACWIP = HaarACWIP)

\end{CodeInput}
\end{CodeChunk}
The arguments of \code{varEWS} are:
\begin{itemize}
  \item \code{object}: The multivariate EWS as a \code{mvLSW} object.
  \item \code{ACWIP}: 4D array containing the autocorrelation wavelet inner product function.
\end{itemize}

Given the variance estimate, the approximate, point-wise 95\% confidence interval for the trivariate EWS elements is derived using \code{ApxCI}. 
\begin{CodeChunk}
\begin{CodeInput}
R> CI <- ApxCI(object = EWS, var = VAR, alpha = 0.05)
\end{CodeInput}
\end{CodeChunk}

The arguments required for the \code{ApxCI} function are:
\begin{itemize}
  \item \code{object}: The multivariate EWS as a \code{mvLSW} object.
  \item \code{var}: Variance estimate of the multivariate EWS as a \code{mvLSW} object.    
  \item \code{alpha}: Type I error.
\end{itemize}

\begin{figure}
  \centering
  \begin{minipage}{.48\textwidth}
    \centering
    \includegraphics[width=\textwidth]{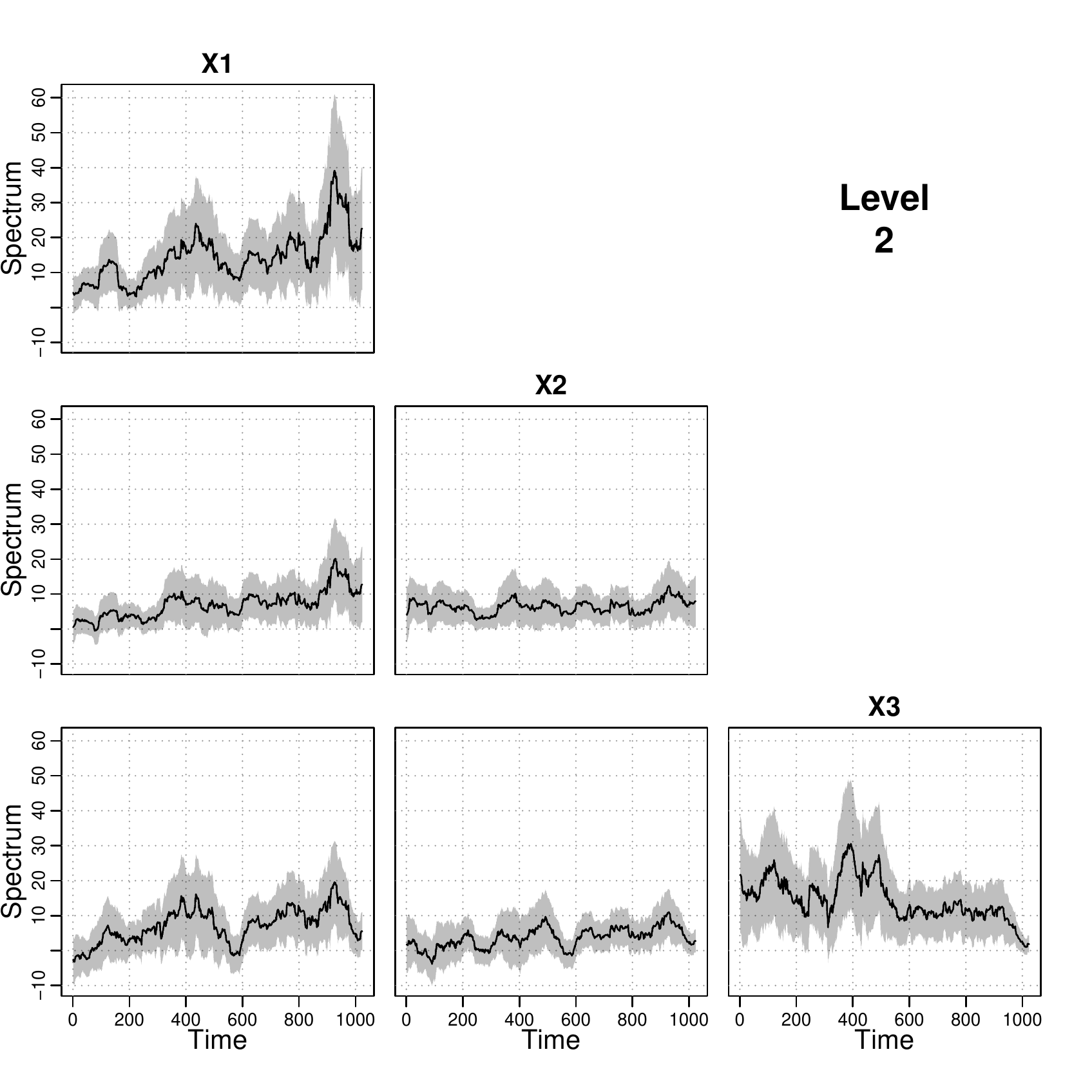}
    \captionof{figure}{Panel plot of the trivariate EWS estimate with approximate, point-wise 95\% confidence intervals at the second level for the simulated time series with Gaussian innervations.}
    \label{fig:Sample_EWS}
  \end{minipage}
  \hfill
  \begin{minipage}{.48\textwidth}
    \centering
    \includegraphics[width=\textwidth]{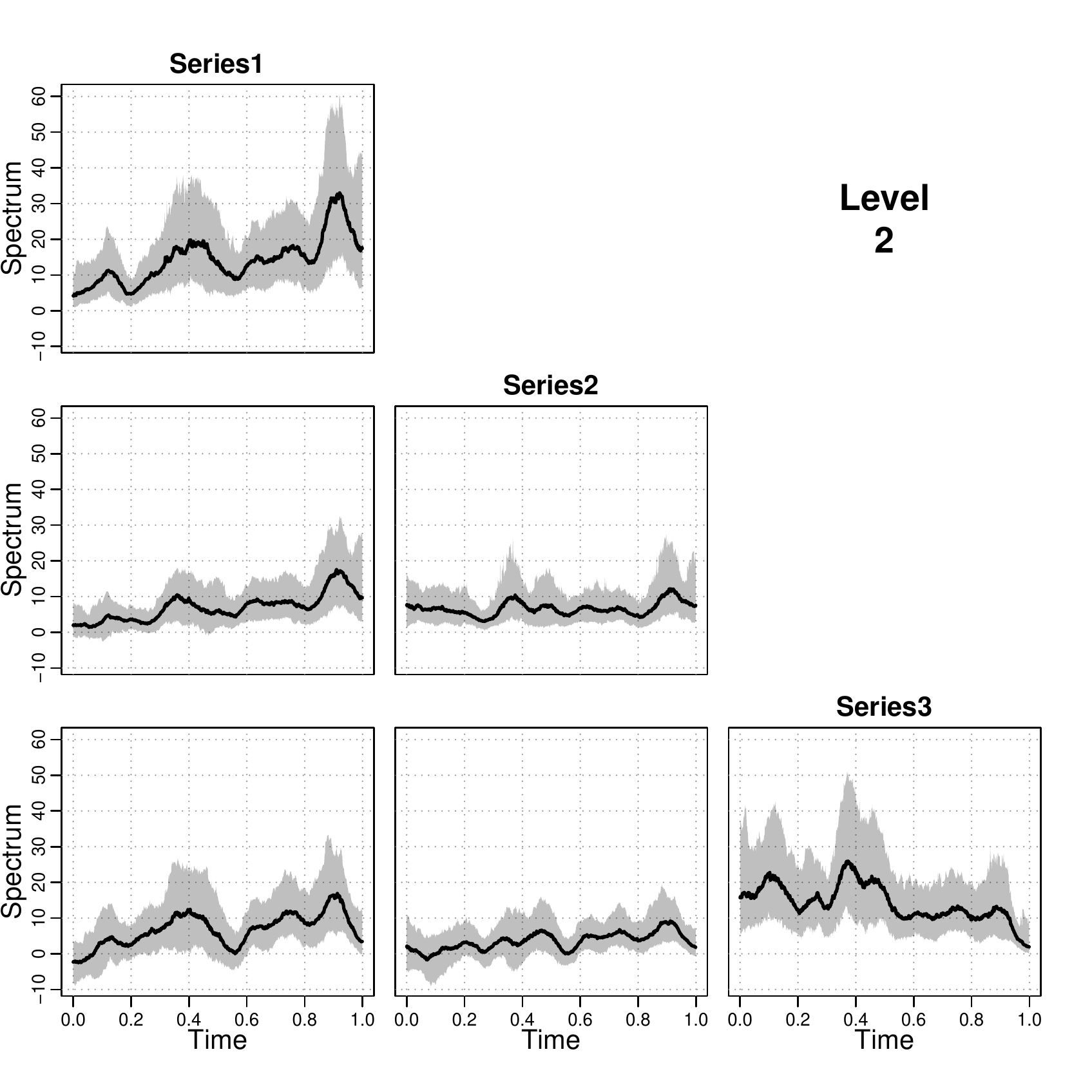}
    \captionof{figure}{Panel plot of the bootstrapped median and central, point-wise 95\% intervals for the trivariate EWS estimate at the second level for the simulated time series with Gaussian innervations.}
    \label{fig:Sample_Boot}
  \end{minipage}
\end{figure}

The returned object is a list with two items named \code{"L"} and \code{"U"}. These are objects having class \code{mvLSW}, that contain the point-wise lower and upper interval bounds respectively. The point-wise confidence intervals can be included on a plot of the multivariate EWS estimate using the \code{plot} command demonstrated in Section~\ref{sec:EG} with the additional \code{Interval} argument. The 95\% confidence intervals for the trivariate EWS estimate in Figure~\ref{fig:Sample_EWS} is generated by the following command:

\begin{CodeChunk}
\begin{CodeInput}
R> plot(x = EWS, style = 2, info = 2, Interval = CI)
\end{CodeInput}
\end{CodeChunk}

To demonstrate the accuracy of the 95\% confidence intervals, the estimates are compared against the 95\% bootstrap intervals. The following extract evaluates the bootstrap intervals from 100 simulated time series with power spectrum defined by the estimate trivariate EWS in Section~\ref{sec:EWS_EG}. The interval estimates for the second level are presented in Figure~\ref{fig:Sample_Boot}. It is clear from comparing these estimates that the analytically derived intervals are close to those obtained via bootstrapping.

\begin{CodeChunk}
\begin{CodeInput}
R> Bsamp <- array(NA, dim = c(P, P, J, T, 100))
R> for(b in 1:100){
R>   X_boot <- rmvLSW(Spectrum = EWS, noiseFN = rnorm)
R>   EWS_boot <- mvEWS(X = X_boot, filter.number = 1, 
+      family = "DaubExPhase", kernel.name = "daniell", 
+      kernel.param = sqrt(T), bias.correct = TRUE)
R>   Bsamp[, , , , b] <- EWS_boot$spectrum
R> }

R> Bl <- as.mvLSW(x = apply(Bsamp, 1:4, quantile, prob = 0.025))
R> Bm <- as.mvLSW(x = apply(Bsamp, 1:4, quantile, prob = 0.5))
R> Bu <- as.mvLSW(x = apply(Bsamp, 1:4, quantile, prob = 0.975))
R> BInt <- list(L = Bl, U = Bu)
R> plot(Bm, style = 2, info = 2, Interval = BInt, lwd = 2)
\end{CodeInput}
\end{CodeChunk}  

\section[European financial indices]{European financial indices}\label{sec:CS_EU}

\begin{figure}
  \centering
  \includegraphics[width=0.9\textwidth]{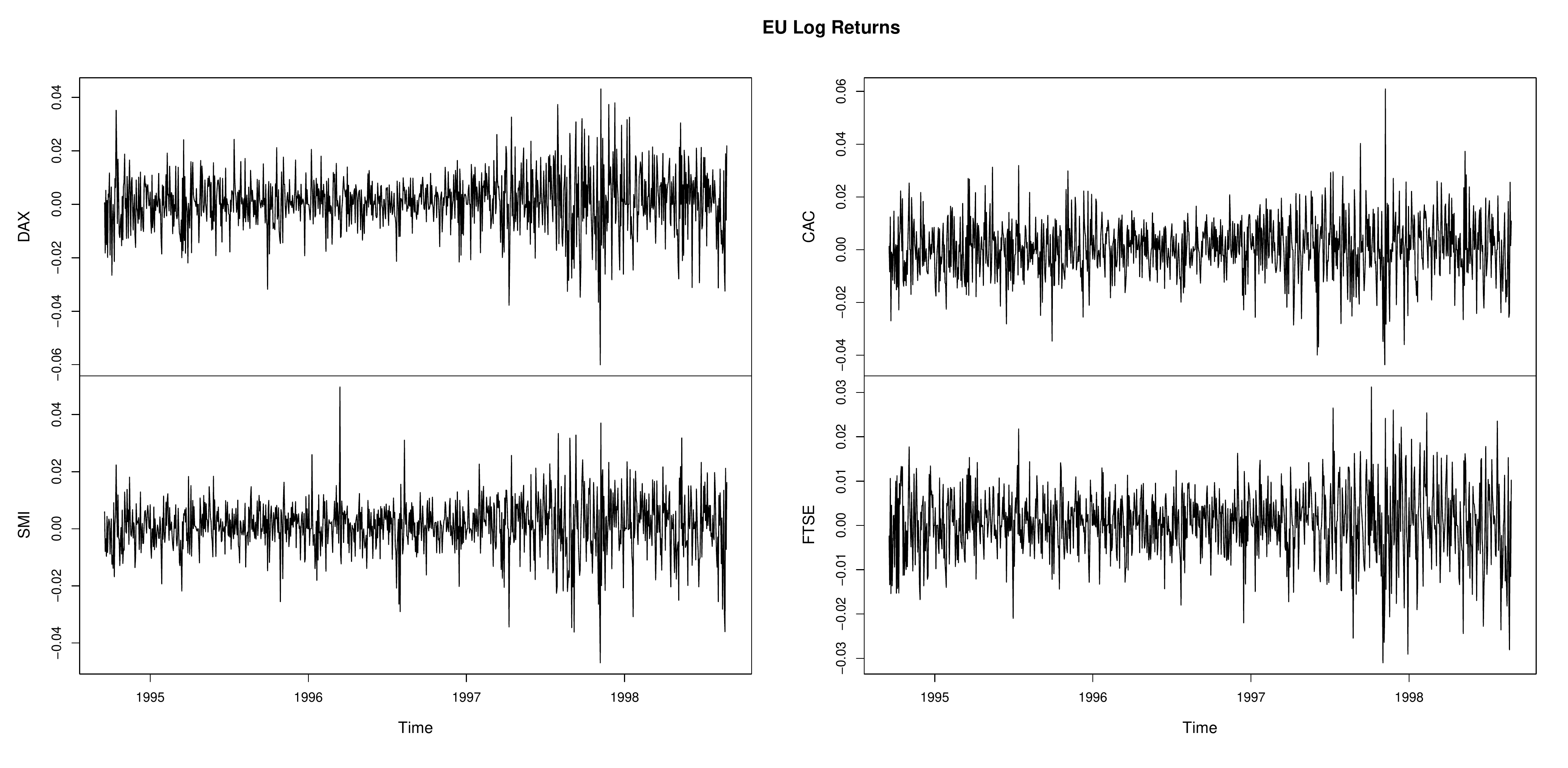}
  \caption{Log-returns time series of four European financial markets. Top-left: German DAX. Top-right: French CAC. Bottom-left: Swiss SMI. Bottom-right: British FTSE.}
  \label{fig:LogReturns_TimeSeries}
\end{figure}

We now illustrate the application of the package using \code{EuStockMarkets}, the European financial indices from the \pkg{datasets} package. The data set contains the daily index of the German (DAX), Swiss (SMI), French (CAC) and British (FTSE) markets. The log-returns are evaluated to remove any trend features and the series is truncated to produce a time series of length $T = 1024$. Figure~\ref{fig:LogReturns_TimeSeries} presents a trace plot of the time series. Each channel of the time series is independently deemed to be non-stationary according to the test implements by \code{BootTOS} from the \pkg{costat} package \citep{Nas13_Rcostat}.
\begin{CodeChunk}
\begin{CodeInput}
R> data("EuStockMarkets", package = "datasets")
R> T <- 1024; J <- log2(T); N <- nrow(EuStockMarkets)
R> EU.lret <- diff(log(EuStockMarkets))
R> EU.lret <- window(EU.lret, start = c(1994, 186))
R> plot(x = EU.lret, main = "EU Log Returns", nc = 2)

\end{CodeInput}
\end{CodeChunk}

\begin{figure}
 \centering
 \includegraphics[width=0.49\textwidth]{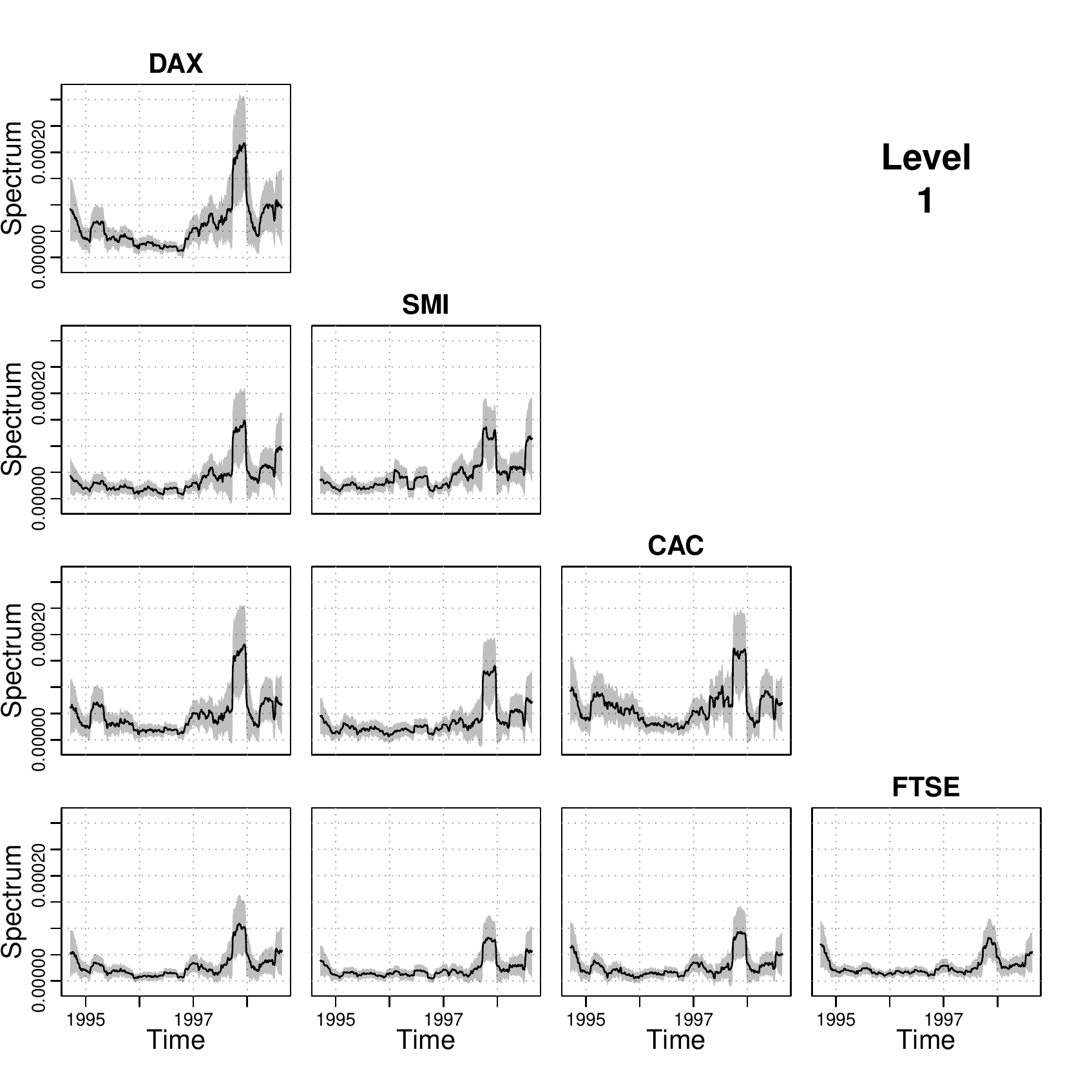}
 \includegraphics[width=0.49\textwidth]{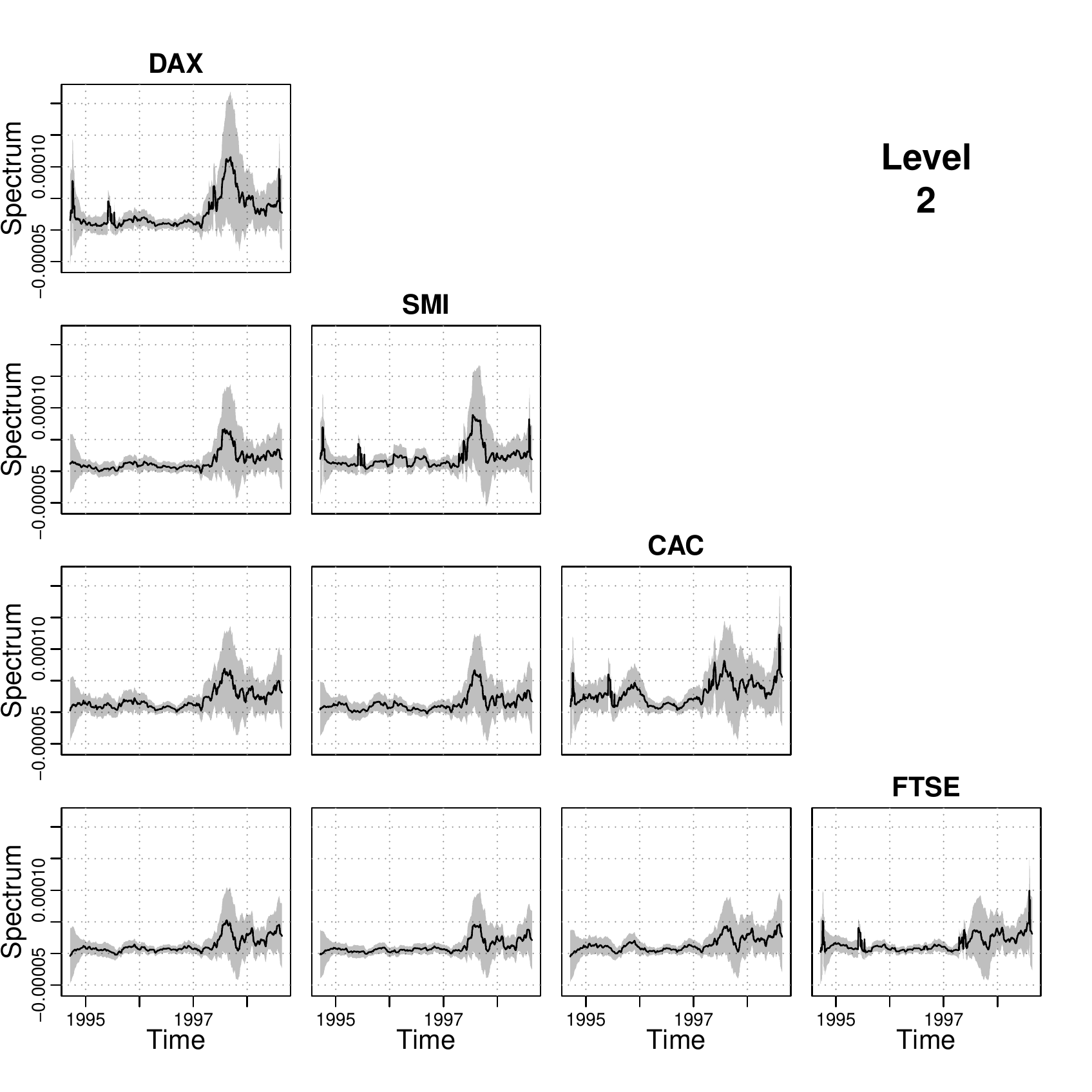}
 \caption{Panel plots at the finest (left) and second (right) levels of the multivariate EWS estimate and approximate, point-wise 95\% confidence intervals for the European market log-returns.}
 \label{fig:LogReturn_EWS}
\end{figure}

\begin{figure}
 \centering
 \includegraphics[width=0.49\textwidth]{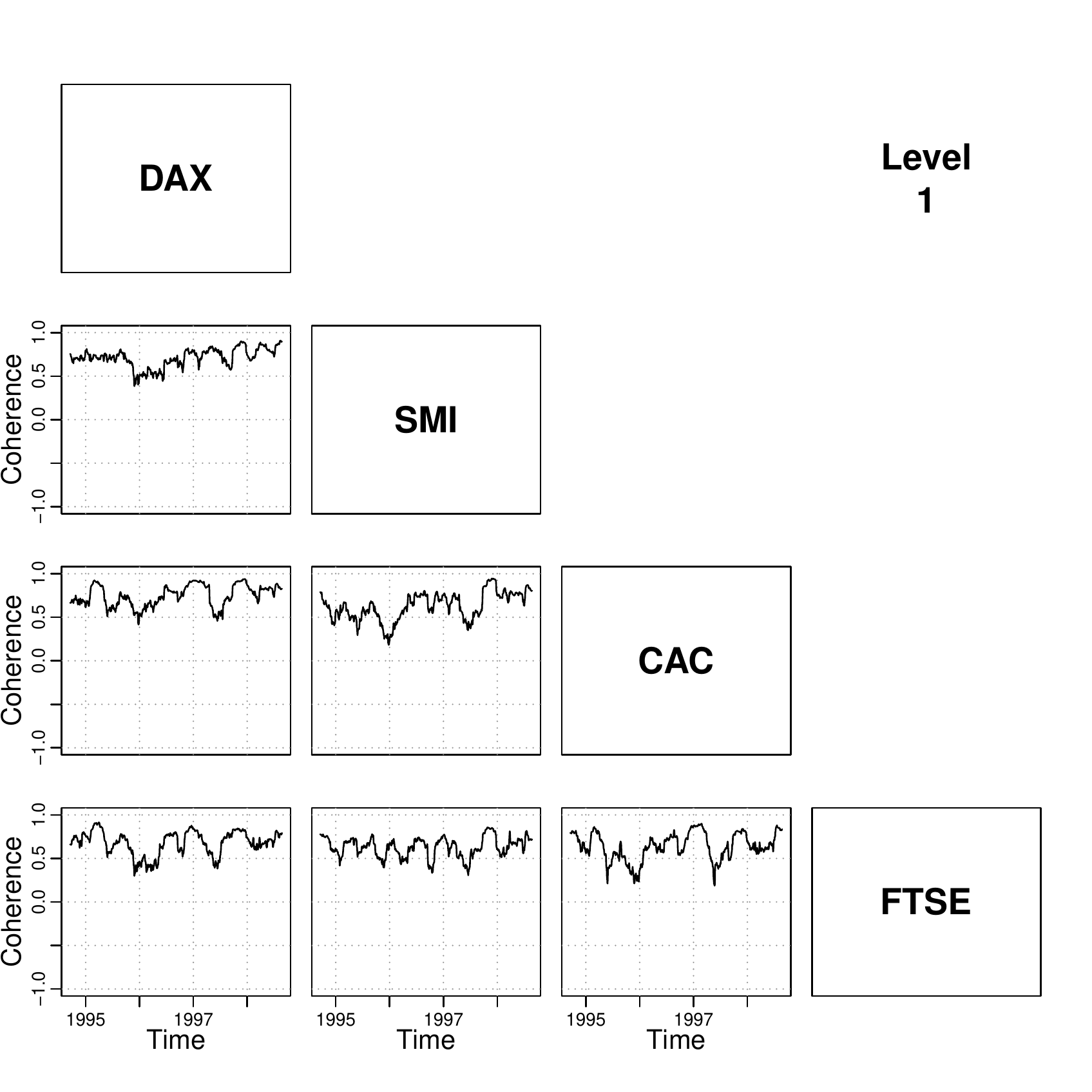}
 \includegraphics[width=0.49\textwidth]{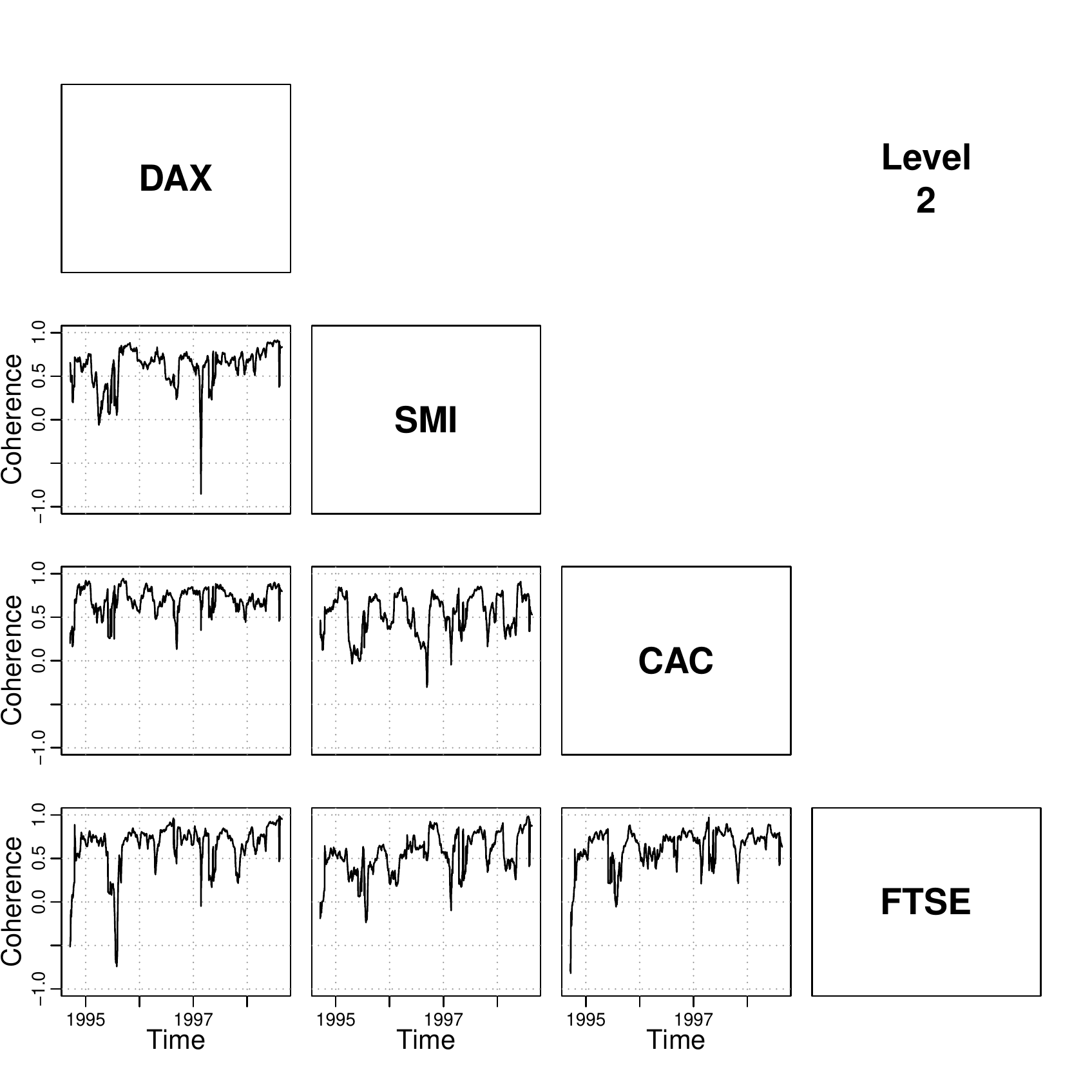}
 \caption{Panel plots at the finest (left) and second (right) levels of the localised coherence estimates for the European market log-returns.}
 \label{fig:LogReturn_RHO}
\end{figure}

\begin{figure}
 \centering
 \includegraphics[width=0.49\textwidth]{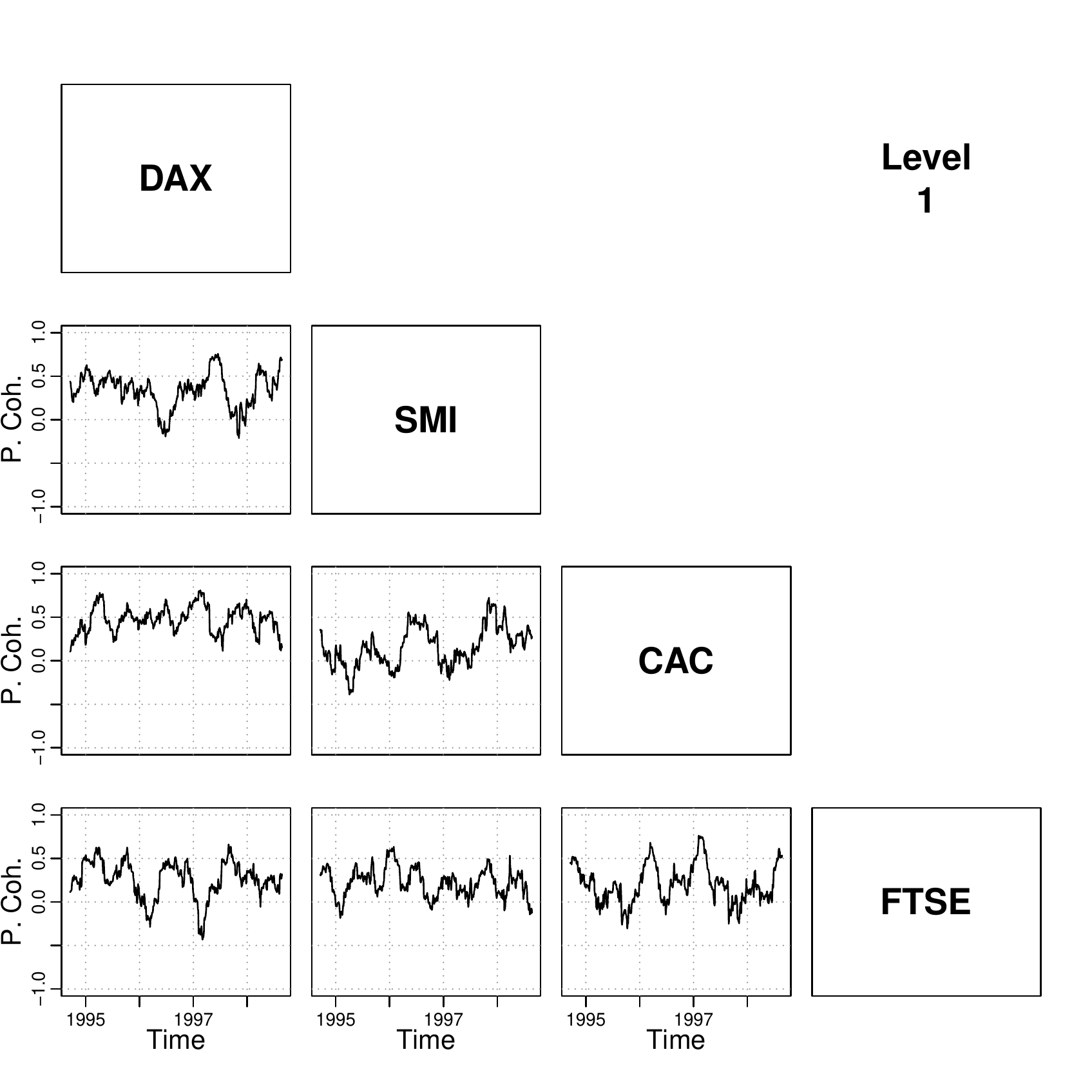}
 \includegraphics[width=0.49\textwidth]{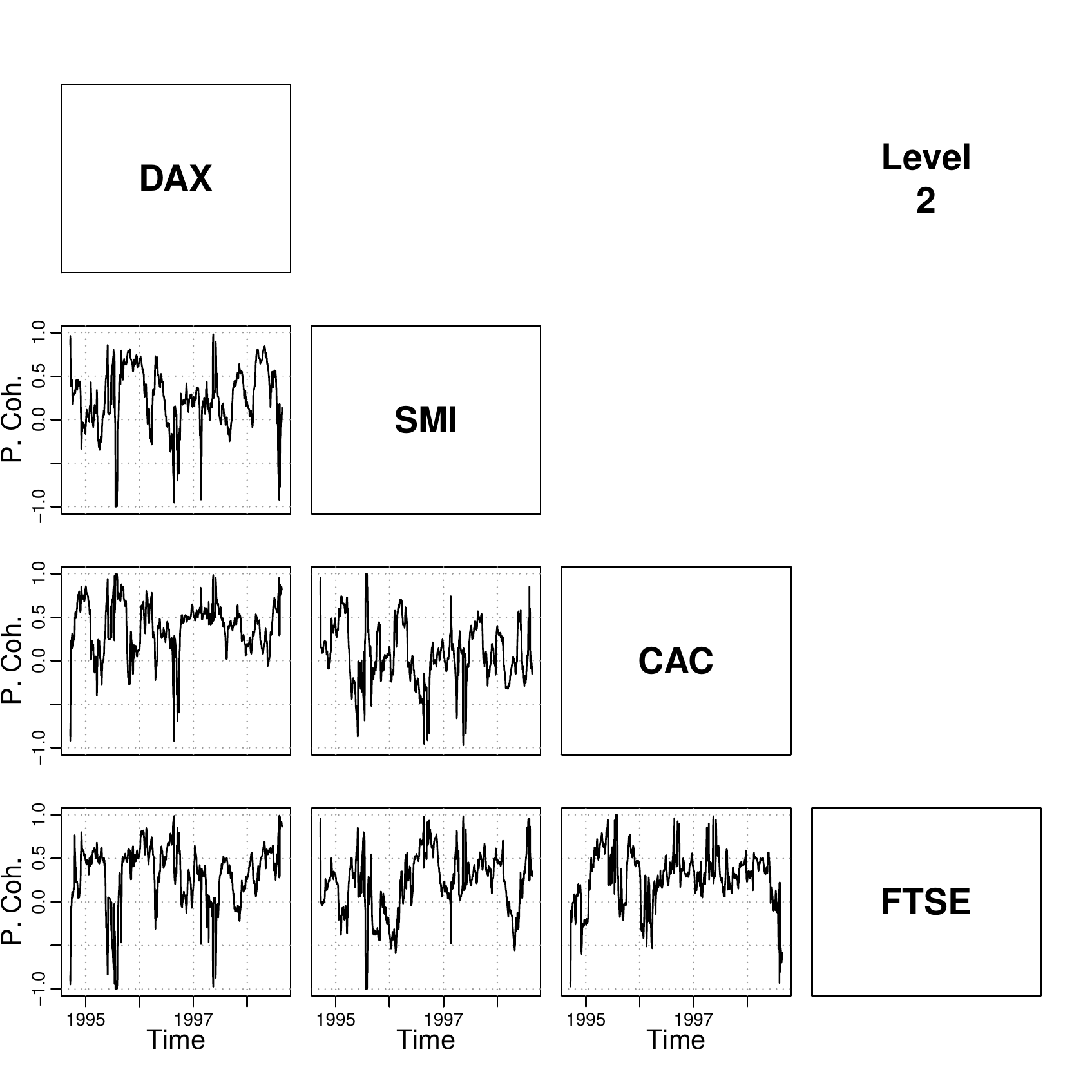}
 \caption{Panel plots at the finest (left) and second (right) levels of the localised partial coherence estimates for the European market log-returns.}
 \label{fig:LogReturn_GAM}
\end{figure}

Estimation of the multivariate EWS is determined using Daubechies extremal phase wavelet with seven vanishing moments, and smoothing is performed with the rectangular kernel function with span parameter $M=\sqrt{T}=32$. The approximate (point-wise) 95\% confidence intervals for the multivariate EWS estimate are evaluated along with the localised coherence and partial coherence estimates. These estimates are obtained using the following \proglang{R} extract, which produces plots of the multivariate EWS, Figure~\ref{fig:LogReturn_EWS}, localised coherence, Figure~\ref{fig:LogReturn_RHO}, and localised partial coherence, Figure~\ref{fig:LogReturn_GAM}; all for the finest and second levels.

\begin{CodeChunk}
\begin{CodeInput}

R> EU.EWS <- mvEWS(X = EU.lret, family = "DaubExPhase", filter.number = 7, 
+    kernel.name = "daniell", kernel.param = sqrt(T), bias.correct = TRUE)
R> ACWIP_FN7 <- AutoCorrIP(J = J, family = "DaubExPhase", filter.number = 7)
R> EU.VAR <- varEWS(object = EU.EWS, ACWIP = ACWIP_FN7)
R> EU.CI <- ApxCI(object = EU.EWS, var = EU.VAR, alpha = 0.05)
R> EU.R <- coherence(object = EU.EWS)
R> EU.G <- coherence(object = EU.EWS, partial = TRUE)

R> plot(x = EU.EWS, style = 2, info = 1, Interval = EU.CI)
R> plot(x = EU.EWS, style = 2, info = 2, Interval = EU.CI)
R> lim <- c(-1, 1); lab1 <- "Coherence"; lab2 <- "P. Coh."
R> plot(x = EU.R, style = 2, info = 1, diag = F, ylim = lim, ylab = lab1)
R> plot(x = EU.R, style = 2, info = 2, diag = F, ylim = lim, ylab = lab1)
R> plot(x = EU.G, style = 2, info = 1, diag = F, ylim = lim, ylab = lab2)
R> plot(x = EU.G, style = 2, info = 2, diag = F, ylim = lim, ylab = lab2)
\end{CodeInput}
\end{CodeChunk}

The non-stationary aspect of the time series is evident in Figure~\ref{fig:LogReturn_EWS} as the spectral power peaks around mid-1997, at both the finest and second levels. The local coherence estimates, Figure~\ref{fig:LogReturn_RHO}, identify that the four channels have strong linear dependence. However, the local partial coherence estimates in Figure~\ref{fig:LogReturn_GAM} are lower. This indicates that some of the linear dependence between a given market pair occurs indirectly because of their relationship with the other (observed) markets. For example, at the finest level, the average coherence between SMI and CAC is 0.63 but its average partial coherence drops to 0.16. The partial coherence estimate at the second and coarser levels becomes more challenging to interpret as some matrices of the multivariate EWS are near singular.

\section[Discussion]{Discussion}\label{sec:discussion}

The \pkg{mvLSW} package contains a number of tools to assess a multivariate non-stationary time series under the locally stationary wavelet framework defined by \citet{Par14}. The main command \code{mvEWS} estimates the multivariate EWS for a LSW time series, Section~\ref{sec:EWS_EG}, and is flexible for the user to specify the best wavelet transform and smoothing kernel that is most appropriate to explore the spectral structure of the time series. However, perhaps more usefully, the package also includes tools for analysing the dependence structure that occurs between pairs of channels, as described by the localised coherence and partial coherence measures presented in Section~\ref{sec:coh_eg}. Uncertainty in the multivariate EWS estimate can be quantified analytically by the routines demonstrated in Section~\ref{sec:CI_eg} in evaluating approximate, point-wise, 95\% confidence intervals for the spectral elements.

A complete case study for examining the spectral structure of the European financial indices \code{EuStockMarkets} was demonstrated in Section~\ref{sec:CS_EU}. This showed that a major non-stationary feature occurs due to a peak in power at the finest level in 1997. Furthermore, analysis of the coherence structure identifies that the indices are strongly and positively linearly dependent with a notable fraction arising from the indirect relationship between all four markets.

\section*{Acknowledgments}

The authors thank Hernando Ombao for many helpful discussions during the early stages of developing this package. They are also gratefully acknowledge Rebecca Killick's contribution of code that implements a regularisation procedure by \citet{Sch99}, which forms part of the \pkg{mvLSW} package.

\bibliography{Multivariate_Locally_Stationary_Wavelet_Process_Analysis}

\end{document}